\documentstyle[aps,prl,preprint,amstex,amssymb]{revtex}

\begin{document}

\draft

\preprint{BARI - TH 169/94, hep-lat/9404017}
\date{April 1994}
\title{
Dual Meissner Effect and String Tension \\ in SU(2) Lattice Gauge
Theory}
\author{Paolo Cea$^{1,2,}$\footnote{Electronic address:
cea@@bari.infn.it} and
Leonardo Cosmai$^{1,}$\footnote{Electronic address:
cosmai@@bari.infn.it}}
\address{
$^1$Dipartimento di Fisica dell'Universit\`a di Bari,
70126 Bari, Italy\\
{\rm and}\\
$^2$Istituto Nazionale di Fisica Nucleare, Sezione di Bari,
70126 Bari, Italy
}
\maketitle
\begin{abstract}
We study the distribution of the color fields due to a static
quark-antiquark pair in SU(2) lattice gauge theory. We find that the
London penetration length measured after Abelian projection in
the Abelian Covariant gauge (Maximal Abelian gauge) agrees with the
one obtained without gauge  fixing. Moreover the penetration length
scales according to asymptotic freedom. We put out a simple relation
between the penetration length and the string tension.
\end{abstract}

\narrowtext

To understand the non~perturbative phenomenon of color confinement in
non~Abelian gauge theories, G.~'t~Hooft~\cite{tHooft75} and
S.~Mandelstam~\cite{Mandelstam76} proposed a model known as dual
superconductor model. The physical grounds of the model lie in the
theory of superconductivity~\cite{Schrieffer92}. In the 't~Hooft's
formulation the dual superconductor model is elaborated in the
framework of the Abelian projection~\cite{tHooft81}. After a
particular gauge has been fixed, reducing the symmetry to that of the
maximal Abelian (Cartan) subgroup, the non~Abelian gauge theory is
described in terms of Abelian projected gauge fields (``photons'').

In this scenario there are also color magnetic monopoles whose
condensation should cause the confinement of all particles which are
color electrically charged with respect to the above photons.
Of course this definition of the color magnetic monopoles does depend
on the Abelian gauge fixing. However the physics, i.e. the monopole
condensation, should be independent on the gauge fixing.

In this paper we analyze the finger-print of the dual superconductor
hypothesis, namely the dual Meissner effect. To do this we analyze the
distribution of the color fields due to a static quark-antiquark pair
in SU(2) lattice gauge theory. Following Ref.~\cite{DiGiacomo90} one
can explore the field configurations produced by the quark-antiquark
pair by measuring the connected correlation function:
\begin{equation}
\label{rhoW}
\rho_W = \frac{\left\langle \mathrm{tr}
\left( W L U_P L^{\dagger} \right)  \right\rangle}
              { \left\langle \mathrm{tr} (W) \right\rangle }
 - \frac{1}{2} \,
\frac{\left\langle \mathrm{tr} (U_P) \mathrm{tr} (W)  \right\rangle}
              { \left\langle \mathrm{tr} (W) \right\rangle } \;\; ,
\end{equation}
where the plaquette $U_P=U_{\mu\nu}(x)$ in the $(\mu,\nu)$ plane is
connected to the Wilson loop $W$ through the Schwinger line L. Note
that the correlation function (\ref{rhoW}) is sensitive to the field
strength rather than to the square of the field
strength~\cite{DiGiacomo93}:
\begin{equation}
\label{rhoWcont}
\rho_W  @>>{a \rightarrow 0}>a^2 g \left[ \left\langle
F_{\mu\nu}\right\rangle_{q\bar{q}} - \left\langle F_{\mu\nu}
\right\rangle_0 \right]  \;.
\end{equation}
According to Eq.(\ref{rhoWcont}) we define the color field strength
as:
\begin{equation}
\label{fieldstrength}
F_{\mu\nu}(x) = \frac{\sqrt{\beta}}{2} \, \rho_W(x)   \;.
\end{equation}
By varying the distance and the orientation of the plaquette $U_P$
with respect to the Wilson loop W, one can scan the color field
distribution of the flux tube.

We performed numerical simulations on $16^4$, $20^4$, and  $24^4$
lattices in the range $2.45 \le \beta \le 2.7$. We used Wilson loop
$L_W \times L_W$, $L_W=L/2 -2$, $L$ being the lattice size.
In order to eliminate the uninteresting short range quantum
fluctuations we cooled our statistical samples. In agreement with
previous studies~\cite{DiGiacomo90,DiGiacomo93}, the connected
correlator $\rho_W$ turns out to be sizeable when the plaquette and
the Wilson loop are parallel. Thus, the component of the
chromoelectric field parallel to the line joining the static charges
is sizeable, while the other components of the chromoelectric field
and the chromomagnetic field are much smaller. Moreover the longitudinal
chromoelectric field is almost constant along the quark-antiquark
line, and it decreases rapidly as the transverse distance is
increased.

Remarkably enough, in a previous study~\cite{Cea93} in the Abelian
covariant gauge (maximal Abelian gauge) we found a similar
behaviour.

In the continuum the Abelian covariant gauge corresponds to impose the
constraints
\begin{equation}
\label{Dmu}
D_\mu A^{\pm}_\mu(x) = 0
\end{equation}
where $A^{\pm}_\mu=A^1_\mu \pm i A^2_\mu$, and $D_\mu$ is the
$A^3_\mu$-covariant derivative. On the lattice the
constraints~(\ref{Dmu}) are implemented by by
maximizing~\cite{Kronfeld87}
\begin{equation}
\label{R}
R= \sum_{x,\mu} \left[ \sigma_3 \widetilde{U}_\mu(x) \sigma_3
\widetilde{U}_\mu^\dagger(x) \right] \;,
\end{equation}
where $\widetilde{U}_\mu(x)$ are the gauge-fixed links:
\begin{equation}
\label{Utilde}
\widetilde{U}_\mu(x) = V(x) U_\mu(x) V^\dagger(x+ \hat{\mu}) \;.
\end{equation}
It is straightforward to check that the residual gauge invariance is
the U(1) group with transformations $g(x) =
{\text{exp}}\left[i\sigma_3\theta(x)\right]$.

In the dual superconductor scenario the long range properties of the
gauge system are encoded into the Abelian fields. On the lattice the
Abelian fields are defined through the Abelian projected links
$U^A_\mu(x)$~\cite{Kronfeld87}:
\begin{equation}
\label{UAbelian}
U^A_\mu(x) = {\text{diag}} \left[ e^{i\theta^A_\mu(x)},
e^{-i\theta^A_\mu(x)}\right] \;,\;
\theta^A_\mu(x) = {\text{arg}} \left[ \widetilde{U}_\mu(x) \right]_{11}
\;.
\end{equation}
In Ref.~\cite{Cea93} we considered the Abelian projected correlator
$\rho_W$:
\begin{equation}
\label{rhoWab}
\rho_W^{A} = \frac{\left\langle \mathrm{tr}
\left(  W^A  U_P^A \right) \right\rangle}
              { \left\langle \mathrm{tr} \left( W^A  \right)
\right\rangle }
 - \frac{1}{2} \,
\frac{\left\langle \mathrm{tr}
\left( U_P^A \right) \mathrm{tr} \left( W^A  \right) \right\rangle}
              { \left\langle \mathrm{tr}
\left( W^A \right) \right\rangle } \; .
\end{equation}
We found that the longitudinal chromoelectric field in the middle of
the flux tube can be fitted according to
\begin{equation}
\label{E_l}
E_l(x_\perp) = \frac{\Phi}{2 \pi}  \frac{1}{\lambda^2}
K_0 \left( \frac{x_\perp}{\lambda} \right) \;,
\;\; x_\perp > 0
\end{equation}
where $x_\perp$ is the transverse distance. $K_0$ is the modified
Bessel function of order zero, $\Phi$ is the external flux, and
$\lambda$ is the London penetration length.

Equation~(\ref{E_l}) is a straightforward consequence of the dual
superconductor hypothesis. Indeed, the transverse shape of the
longitudinal chromoelectric field is the dual version of the Abrikosov
vortex field distribution. However, the chromoelectric field
has been obtained through the
gauge-dependent correlator $\rho^A_W$. So the physical meaning of
Eq.~(\ref{E_l}) is not clear. On the other hand, if the confinement is
realized by the dual superconductor mechanism, then Eq.(\ref{E_l})
should hold even if we use the gauge invariant correlator $\rho_W$.
To check this point,
we measured $E_l(x_\perp)$ in the middle of the flux tube
both in the Abelian covariant gauge and without gauge fixing by
varying $x_\perp$ up to transverse distance $L_W$.

Remarkably enough, we find that Eq.~(\ref{E_l}) holds for both
definition of the longitudinal chromoelectric
field~\cite{preliminary}. We obtain a rather good fit to the data
($\chi^2/{\text{d.o.f}} \lesssim 1$) if $x_\perp \ge 2$ (in lattice
units); this is reasonable since Eq.~(\ref{E_l}) applies in the region
$\lambda \gg \xi$, where $\xi$ is the coherence length (more on this
later on).

In Figure~1 we report our results for the inverse of the penetration
length $\mu = 1/\lambda$.
For the gauge invariant correlator $\rho_W$ we collected up to 100
configurations with 8 cooling steps (we, however, checked that the
results are stable up to 12 cooling steps) at each $\beta$. We
performed measurements for 9 different values of $\beta$ in the range
$2.45 \le \beta \le 2.7$ (for figure readability not all the values
are displayed).

In the Abelian covariant gauge we performed 100-500 measurements for
7 different values of $\beta$ in the same range. The gauge is fixed
iteratively via the overrelaxation algorithm of Ref.~\cite{Mandula87}
with the overrelaxation parameter $\omega=1.7$.

Figure 1 suggests that both penetration lengths scale according to
two-loop asymptotic scaling law. We find
$\mu/\Lambda_{\overline{MS}}=8.9(3)$ ($\chi^2/{\text{d.o.f.}} \simeq
2$) for the gauge invariant correlator, and
$\mu/\Lambda_{\overline{MS}}=8.3(7)$ ($\chi^2/{\text{d.o.f.}} \simeq
0.4$) in the Abelian covariant gauge. A striking consequence of Fig.1
is that both definitions of the penetration length agree. The overall
fit of the data gives (dashed lines)
\begin{equation}
\label{mu_lambda}
\frac{\mu}{\Lambda_{\overline{MS}}} = 8.8(3)
\end{equation}
with $\chi^2/{\text{d.o.f.}} \simeq 1.8$. Thus our results strongly
suggests that the London penetration length is a physical
gauge-invariant quantity.

In Figure~2 we display the external flux $\Phi$ versus $\beta$.
Naively we expect that
\begin{equation}
\label{Phi}
\Phi = 1  \;.
\end{equation}
Indeed in the Abelian covariant gauge, we see that the data are quite
close to Eq.~(\ref{Phi}). On the other hand, without gauge fixing the
external flux is strongly affected by lattice artefacts which seem to
disappear by increasing $\beta$. Thus we are led to suspect that the
external flux gets renormalized by irrelevant operators, whose effects
are strongly reduced in the Abelian covariant gauge. As we will argue
in a moment, these effects are responsible for the non-scaling
behaviour of the string tension. Indeed, the string tension can be
defined as the energy stored into the flux tube per unit length.
By observing that the longitudinal
chromoelectric field is almost constant along the flux tube, we obtain
from Eq.~(\ref{E_l}):
\begin{equation}
\label{string}
\sqrt{\sigma} = \frac{\Phi}{\sqrt{8 \pi}} \mu \;.
\end{equation}
Now, we have shown that $\mu$ almost scales according to asymptotic
freedom. Whence the non-scaling behaviour of the string tension is due
to the lattice renormalization of the external charge. However,
Eq.~(\ref{string}) allows us to get rid of these effects by putting
$\Phi=1$ into Eq.~(\ref{string}):
\begin{equation}
\label{string_Phi}
\sqrt{\sigma} = \frac{\mu}{\sqrt{8 \pi}} \;.
\end{equation}
In Figure~3 we show Eq.~(\ref{string_Phi}). For comparison we display
also the string tension extracted from the Wilson loops (full points).
Due to low statistics and limited lattice size our extimation of the
string tension is poor. However, it is gratifying to find that the
linear extrapolation to the continuum limit of our data is compatible,
albeit with large error, with the one obtained on larger
lattices~\cite{Fingberg93} (full star):
\begin{equation}
\label{string_Wilson}
\frac{\sqrt{\sigma}}{\Lambda_{\overline{MS}}} = 1.79 \pm 0.12 \;.
\end{equation}
The string tension defined by Eq.~(\ref{string_Phi}) (open points)
seems to scale for $\beta \ge 2.45$ to a value which is consistent
with (\ref{string_Wilson}). Indeed we find (dashed lines)
\begin{equation}
\label{string_flux}
\frac{\sqrt{\sigma}}{\Lambda_{\overline{MS}}} = 1.76 \pm 0.15 \;,
\end{equation}
where the quoted error include our extimation of the systematic error.

We would like to comment on the validity of Eq.~(\ref{string_Phi}). In
order to obtain Eq.~(\ref{string_Phi}) we extrapolated Eq.~(\ref{E_l})
up to $x_\perp = 0$. This extrapolation leads to a negligible error in
(\ref{string_Phi}) if $\lambda/\xi \gtrsim 2$. So it is important to
have an extimation of $\xi$. The coherence length is determined by
the monopole condensate, the order parameter for the confinement.
Unfortunately we do not yet have at disposal a good definition of the
order parameter.\\
Recently two different groups~\cite{Singh93,Matsubara93} give an
extimation of the coherence length by using monopole currents defined
following DeGrand and Toussaint~\cite{DeGrand80}. However we feel that
one can not rely on this approach. As a matter of fact, the authors of
Ref.~\cite{DelDebbio91} argued convincingly that the DeGrand-Toussaint
monopole density is not the order parameter for the confinement.

Let us conclude by stressing the main results of the paper. We found
that the London penetration length in the dual Meissner effect is a
physical quantity. Moreover we put out a very simple relation between
the string tension and the penetration length. A remarkable
consequence of our Eq.~(\ref{E_l}) is that the long range properties
the SU(2) confining vacuum can be described by an effective Abelian
theory. In addition, after fixing the gauge with the
constraints~(\ref{Dmu}), it seems that the degrees of freedom which
are not relevant to the confinement get suppressed. In this respect
the situation looks like the Bardeen-Cooper-Schrieffer (BCS) theory of
superconductivity. As it is well known~\cite{Schrieffer92}, in the BCS
theory one considers a reduced Hamiltonian which, however, breaks the
electromagnetic gauge invariance. Nevertheless, the reduced BCS
Hamiltonian, by retaining the degrees of freedom relevant to the
superconductivity, gives the correct explanation of the Meissner
effect.

Let us conclude by stressing that that the most urgent problem to be
addressed in the future studies is the reliable extimation of the
coherence length~\cite{DiGiacomo94}.
%
%

%
%
%
\begin{figure}
\label{Fig1}
\caption{The inverse of the penetration length in units of
$\Lambda_{\overline{MS}}$ as a function of
$a\Lambda_{\overline{MS}}$. Unfixed gauge: circles $L=16$, triangles
$L=20$,
square $L=24$. Abelian covariant gauge: crosses $L=16$,
diamond $L=20$.}
\end{figure}
\begin{figure}
\label{Fig2}
{\caption{The external flux versus $\beta$. Symbols as in Fig.1.}}
\end{figure}
\begin{figure}
\label{Fig3}
\caption{The square root of the string tension in units of
$\Lambda_{\overline{MS}}$ versus $a\Lambda_{\overline{MS}}$.
Open symbols refer to
Eq.~({\protect{\ref{string_Phi}}}).
Full star is the extrapolated continuum limit.}
\end{figure}
%
%
%
\end{document}